\documentclass[superscriptaddress,aip,rsi,amsmath,amssymb,floatfix,10pt,twocolumn]{revtex4-1}

\usepackage{braket}
\usepackage{graphicx}
\usepackage{float}
\usepackage{hyperref}
\usepackage{amsmath}
\usepackage{mathrsfs}
\usepackage[dvipsnames]{xcolor}
\usepackage{longtable}
\usepackage{amsfonts}
\usepackage{tikz}
\usepackage[percent]{overpic} 
\usepackage{bm}
\usepackage{xcolor}
\usepackage[most]{tcolorbox}
\usepackage{CJK}
\usepackage{cancel}

\definecolor{arrowcolor}{RGB}{10,10,100}

\usepackage{mdframed}
\usepackage{appendix}

\usepackage{nicefrac}

\newcommand{\Applied}{Department of Applied Physics, Stanford University, Stanford, CA}
\newcommand{\Physics}{Department of Physics, Stanford University, Stanford, CA}

\usepackage{dcolumn}
\usepackage[utf8]{inputenc}
\usepackage[T1]{fontenc}
\usepackage{mathptmx}
\usepackage{etoolbox}

\makeatletter
\def\@email#1#2{%
 \endgroup
 \patchcmd{\titleblock@produce}
  {\frontmatter@RRAPformat}
  {\frontmatter@RRAPformat{\produce@RRAP{*#1\href{mailto:#2}{#2}}}\frontmatter@RRAPformat}
  {}{}
}%
\makeatother

\begin{document}
\title{A Second-Order Optical Butterworth Fabry-P\'erot Filter}
\author{Zeyang Li}\thanks{These authors contributed equally.}
\affiliation{\Applied}
\author{Abhishek V. Karve}\thanks{These authors contributed equally.}
\affiliation{\Applied}
\author{Xin Wei}\thanks{These authors contributed equally.}
\affiliation{\Applied}
\author{Jonathan Simon}\email{jonsimon@stanford.edu}
\affiliation{\Applied}
\affiliation{\Physics}
\begin{abstract}
    Filters with flat-top pass-bands are a key enabling technology for signal processing. From communication to sensing, the ability to choose a pass \emph{band}, rather than a single pass \emph{frequency}, while still efficiently suppressing backgrounds at other frequencies, is a critical capability for ensuring both detection sensitivity and power efficiency. 
Efficient transmission of a single frequency can be achieved by a single-pole resonator---which in optics is a Fabry-P\'erot cavity offering linewidths from kHz to GHz and beyond. Coupling multiple resonators allows for the construction of flat-top multi-pole filters. These, although straightforward from RF to THz where resonators are macroscopic and tunable, are more difficult to control in the optical band and typically realized with dielectric stacks, whose passband widths exceed 100 GHz. Here, we bridge the gap to narrower bandwidth flat-top filters by proposing and implementing a second-order Butterworth-type optical filter in a single two-mirror Fabry-P\'erot cavity, by coupling the two polarization modes. We demonstrate a pass-band width of 2.68(1)~GHz, a maximum stopband suppression of 43~dB, and a passband insertion loss of 2.2(1)~dB, with out-of-band power suppression falling as the fourth power of detuning. This approach is viable down to much narrower filters, and has the potential to improve high-frequency phase noise performance of lasers, enhance the sensitivity of LIDARs, and provide higher quality narrowband filtering, for example, for Raman spectroscopy. 
\end{abstract}
\maketitle
\section{Introduction}

Band-pass filters (with a flat passband) are a standard tool in modern microwave and RF engineering to separate a desired signal from undesirable components at other frequencies.  In the optical regime, dielectric-stack filters can achieve similar effects~\cite{johnson2016bandpass,quarterwave2025,bandpassFuture}, but with a slope's width at the 0.1~nm ($\sim 50\;$GHz at 780~nm) level. Such filters are not ideal for certain precise applications, including many in atomic physics experiments, where the required passband is typically no more than a few GHz wide~\cite{Kimble2021PRDFilterCavityLIGO,Levine2018PRLHighFidelity,Li2022PRApplied,Hald:05,Ludlow:21}.

On the other hand, optical cavities can serve as narrow first-order filters~\cite{readoutModeCleaner}, efficiently transmitting a single frequency before \emph{immediately} suppressing signals off of the transmission resonance, with a field-suppression that scales inversely with detuning, with bandwidths down to kHz~\cite{1hzFP,10khzFP}. Fabry-P\'erot filters, for example, are commonly employed for frequency-selection and stabilization in lasers~\cite{Kessler2012-sb,PhysRevA.77.053809,PhysRevA.107.042611,PhysRevA.111.042614}, Raman spectroscopy~\cite{Hoffman:13, EtalonRamanSpec1,EtalonRamanSpec2,Levine2022PRA}, sum frequency generation spectroscopy~\cite{SFG1,SFG2}, double resonance 2D-IR spectroscopy~\cite{2dIR1,2dIR2}, and pulse-laser shaping~\cite{ChenEtalonShaping,Hoffman:13}. Such a cavity, however, is a first-order filter with a Lorentzian shape, characterized by the slow $1/\delta^2$ roll-off with the laser detuning from the filter center ($\delta$) in the wings without a flat transmission region at the top. As an example, distributed Bragg reflector (DBR) lasers have been shown to stabilize to subkilohertz linewdiths by using long external cavities~\cite{BraggLaser2012}. This approach would benefit from a filter that has a flat-top transmission with faster roll-off in its wings.

High-order---such as Butterworth-type---filters~\cite{butterworth1930theory}, can achieve a flatter transmission band and a faster roll-off within the rejection band. Efforts towards flat-top optical filters have been primarily aimed at wideband nano- / micro-photonic~\cite{scheuer2013trap,Xu2019PRA} devices, where the small size and relatively low finesse of the filters combine to ensure that the filter passbands never reach the GHz-to-sub GHz regime. Active gain media have been proposed as an ingredient of Parity-Time (PT)-symmetric flat-top filters~\cite{Habler2020PRA}, but the fine-tuned nature of the device makes it generally difficult to realize, and its added noise makes it fundamentally incompatible with filtering of single photon signals.

Realizing such narrowband, high-order optical filters should be possible by cascading Fabry P\'erot resonators~\cite{van1985multimirror,nanoCoupledCavs}, and while coupled macroscopic systems have been controlled~\cite{stone2020optical}, simultaneously locking multiple resonators is technically challenging. In this letter, we report our demonstration of a compact \& robust free-space second-order Butterworth optical filter where the two employed resonator modes are \emph{geometrically frequency locked to one another}, achieved by coupling the two polarization modes of a single two-mirror cavity.

\section{Second order filters in terms of coupled resonators}

\begin{figure}
    \centering
    \includegraphics[width=.9\linewidth]{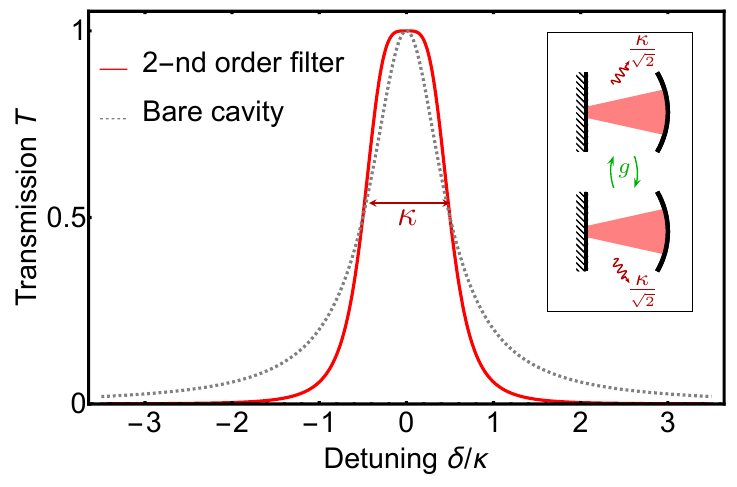}
    \caption{Theoretical spectra of the second-order filter. The transfer function $T$ (red solid line) of the second-order coupled cavity system is shown as a function of detuning between the frequency of the input light and the central resonant frequency of the system. This is compared with the bare cavity Lorentzian transmission (gray dashed line). The two transfer function spectra share the same peak transmission and a full width at half maximum (FWHM) of $\kappa$, while the second-order filter exhibits a flatter top and faster roll-off in its wings. The inset shows how such a filter can be thought of as two coupled single-ended cavity resonators, each with a power loss rate of $\nicefrac{\kappa}{\sqrt{2}}$. The coupling Rabi frequency between the two cavity modes is defined as g. To achieve a flat-top transmission spectrum, the coupling needs to be tuned to $g = \nicefrac{\kappa}{\sqrt{2}}$, as explained further in the main text.}
    \label{fig:FilterTransfer}
\end{figure}

A second-order filter with a full width at half maximum (FWHM) of $\kappa$---can also be thought of as the power loss rate---can be implemented in transmission through two cascaded cavities, each with a FWHM (due to their non-shared mirrors) of $\nicefrac{\kappa}{\sqrt{2}}$, and a coupling Rabi frequency of $\nicefrac{\kappa}{\sqrt{2}}$ as well~\cite{Naaman2022,MYJbook1980,yan2023broadband}, as defined in the inset in Fig.~\ref{fig:FilterTransfer}. This precise matching between the input/output rate and the resonator coupling strength can be thought of as a kind of optical impedance matching that ensures \emph{both} a flat-topped spectral response \emph{and} $100\%$ transmission on resonance.

This device acts as a second-order filter because the light must pass through \emph{two} cavities, in sequence, to be transmitted; the top of the filter is flattened by mode splitting induced by the coupling between the cavities. Viewed another way: the mode coupling generates two ``dressed'' modes, each a superposition of excitations in both cavities. The transmitted fields from the two modes interfere constructively \emph{between} the modes, leading to a flat-topped transmission. The fields interfere destructively \emph{outside} of the two modes, leading to the second-order field suppression.

This approach is fundamentally different from having a cavity followed by an optical isolator followed by a cavity. Such an object would act as two first-order filter and have a transmission spectrum of a Lorentzian squared. Such a response also provides the faster roll-off in the wings but does not provide the desired flat-top transmission. Allowing for the back coupling of the second cavity into the first cavity is what gives the flat-top \emph{and} the faster roll-off in the wings.

The non-Hermitian Hamiltonian that describes the coupled-cavity system that yields an ideal second-order filter is~\cite{MYJbook1980}:
\begin{align}\label{eq:ButterworthHamiltonian}
    \hat{H}_\mathrm{2nd} = \omega_c\mathbb{I}_2(\hat{a}^\dagger\hat{a}+\hat{b}^\dagger\hat{b})+\frac{1}{2}\left(\frac{\kappa}{\sqrt{2}}\right)
    \begin{matrix}
        (\hat{a}^\dagger & \hat{b}^\dagger)\\\phantom{a}
    \end{matrix} 
    \Bigg(\begin{matrix}
        i&1\\1&i
    \end{matrix} \Bigg)
    \Bigg(\begin{matrix}
        \hat{a}\\\hat{b}
    \end{matrix} \Bigg),
\end{align}
where $\hat{a}$ and $\hat{b}$ are anniliation operators of the two cavity modes, $\mathbb{I}_2$ is the $2 \times 2$ identity, $\omega_c$ is the center resonance frequency of the filter, and $\kappa$ is the full width at half maximum of the filter as defined in Fig.~\ref{fig:FilterTransfer}. Since the Hamiltonian is linear, we will omit the creation and annihilation operators and use the $2\times2$ matrix to represent the Hamiltonian, with the bases $\ket{1}\equiv\hat{a}^\dagger\ket{\mathrm{vac}}$ and $\ket{2}\equiv\hat{b}^\dagger\ket{\mathrm{vac}}$ where $\ket{\mathrm{vac}}$ is the vacuum state of both cavity modes. 
By using the non-Hermitian perturbation theory~\cite{simonThesis, stone2020optical,ResolventMethod}, the ratio of the output power from the second cavity to the input power of the first cavity with laser frequency $\omega_l$ is given by
\begin{align}\label{eq:ButterworthTransmission}
    T = \left|  \frac{\kappa}{\sqrt{2}}
    \left\langle 1 \middle| 
    \frac{1}{\hat{H}_\mathrm{2nd} - \omega_l \mathbb{I}_2} 
    \middle| 2 \right\rangle \right|^2
    = \frac{1}{1 + (2\delta/\kappa)^4},
\end{align}
where $\delta\equiv\omega_l-\omega_c$ is the driving detuning from the cavity. 
The transmission can also be derived from a scattering matrix approach~\cite{stone2020optical}, as elaborated in Appendix A. 

The shape exhibits an ideal second-order flat-top and a fast-decaying transfer function with a FWHM of $\kappa$. The Hamiltonian can be diagonalized into the form,
\begin{align}\label{eq:splittingForm}
    \hat{H}_\mathrm{2nd,d} = \omega_c\mathbb{I}_2+\frac{1}{2}\left(\frac{\kappa}{\sqrt{2}}\right)
\left(\begin{matrix}
        i+1&0\\0&i-1
\end{matrix}\right),
\end{align}
the eigenmodes of which are split by $\nicefrac{\kappa}{\sqrt{2}}$.

\subsection{Polarization-Basis Coupled Cavity}

\setlength{\unitlength}{1\columnwidth}
\begin{figure}[!t]
\includegraphics[width=\columnwidth]{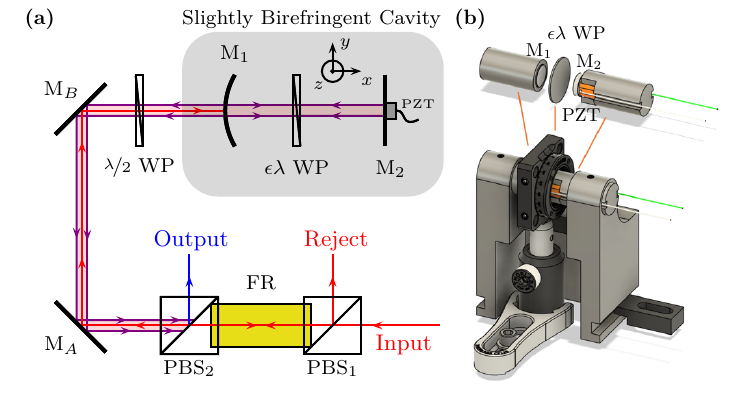}
\caption{Experimental setup of the second-order filter. (a) Polarization basis coupled cavity scheme for the second-order optical filter. The polarization of the input light (red) is aligned such that it transmits through the polarizing beam splitter (PBS$_1$), a Faraday Rotator (FR), and PBS$_2$. The two PBSs and the FR are part of Thorlab's IO-3-780-HP optical isolator. The polarization is then matched to the cavity polarization mode by a $\nicefrac{\lambda}{2}$ wave plate (WP). It is then incident on a curved, partially reflective mirror M$_1$. The coupling of the polarization modes is achieved through an intra-cavity birefringent optic that acts like a $\epsilon \lambda$ WP---two implementations of such an optic are discussed in the main text. The cavity length is stabilized by the piezoelectric stack (PZT) on the high-reflectivity end mirror (M$_2$), by locking on the slope to a different wavelength (5~nm away), which is separated from the output by a conventional dielectric-stack filter. The cavity is single-ended, and all light goes back through M$_1$. The output (purple, shown as a bigger beam for convenience) carries a polarization superposition of the input light and its orthogonal component. The orthogonal polarization (blue) is reflected by PBS$_2$ and is the `output' light of the filter. The input polarization component (red) gets reflected by PBS$_1$ and is the `rejected' light of the filter. (b) CNC-machined monolithic mount of the cavity structure to suppress acoustic noise. The zoomed-in view shows the mirrors and the birefringent optic as they relate to the schematic in (a).
}
\label{fig:Pol2ndScheme}
\end{figure}

There are various ways to implement the coupled cavities; the most direct is to employ three dielectric mirrors in sequence, akin to prior mode-conversion work~\cite{stone2020optical}. Such an approach necessitates precise stabilization of two \emph{separate} resonator lengths, fine-tuned control of the transmissive coupling between the two cavities, and precise mode-matching.

Here, we explore a simpler approach, leveraging the two polarization modes of a \emph{single} two-mirror Fabry-Pérot cavity, with birefringence-induced coupling between the modes. The coupling needed for the Hamiltonian in \eqref{eq:splittingForm}, can---when the two modes under consideration are orthogonal polarization states---arise \emph{directly} from birefringence. Such a coupling can be turned on and controlled via an intra-cavity birefringent optic. This single-cavity implementation benefits us by automatically matching the loss rates of the two cavity modes. This approach provides a system with fewer degrees of freedom that must be actively stabilized, and continuous control of mode coupling through the tunable birefringence.

We explore two approaches to generate a small tunable birefringence in a low-loss (and thus cavity-compatible) optic: The first is to use a high ($m$th-) order half-wave plate (HWP). At normal incidence, the double-pass birefringence is exactly $2\pi$ and thus the modes are unsplit. To induce a controlled mode-coupling, we \emph{tilt} the intra-cavity HWP about the z-axis for a non-normal incidence angle $\theta$–deviating from an ideal HWP, as shown in Fig.~\ref{fig:Pol2ndScheme}(a). We then \emph{rotate} the waveplate such that its extraordinary axis is at an angle $\phi$ from the $z$-axis. The ordinary and extraordinary axes of the HWP can be decomposed into their Cartesian components, 
\begin{align*}
    \hat{e}_o&=\sin\phi\hat{e}_z-\cos\phi(\cos\theta\hat{e}_y-\sin\theta\hat{e}_x),\\
    \hat{e}_e&=\cos\phi\hat{e}_z+\sin\phi(\cos\theta\hat{e}_y-\sin\theta\hat{e}_x).
\end{align*}

For the cavity light in polarization $\hat{e}_\alpha=\cos\alpha\hat{e}_z+\sin\alpha\hat{e}_y$, the round-trip shift is
\[\begin{split}
    \Delta\Phi(\alpha)=&\frac{2(m+1)\pi}{\cos\theta}|\hat{e}_\alpha\cdot\hat{e}_e|\\
    =&\frac{2(m+1)\pi}{\cos\theta}(\sin\phi\cos\theta\sin\alpha+\cos\phi\cos\alpha),
\end{split}
\]
which corresponds to the two eigenmodes at $\alpha=\arctan(\cos\theta\tan\phi),\; \arctan(\cos\theta\tan\phi)+\pi/2$ with a frequency splitting of,
\begin{align}
\frac{2g}{2\pi\times\mathrm{FSR}}=\sqrt{1-\sin^2\theta\sin^2\phi}\times(m+1)\frac{1-\cos\theta}{\cos\theta},
\end{align}
where FSR stands for the free spectral range. This indicates that the coupling strength $g$ is coarsely tuned by $\theta$ and finely-tuned by $\phi$, which also guarantees a wide tuning range of $g$, from zero to $2\pi\times\mathrm{FSR}$. 

Our second approach leverages the fact that, in practice, the required birefringence is quite low. We introduce stress birefringence into the cavity by applying a uniaxial force to an otherwise non-birefringent thin glass plate. This approach avoids having to tilt any intra-cavity optics and lowers the minimum cavity length–--up to considerations of the radius of curvature of the mirrors–--to the thickness of the glass plate ($\sim 1$~mm), enabling larger resonator free-spectral ranges up to $\sim 300$~GHz. Data in Fig.~\ref{fig:TransmissionData} were collected using this approach.

\begin{figure}
\includegraphics[width=\columnwidth]{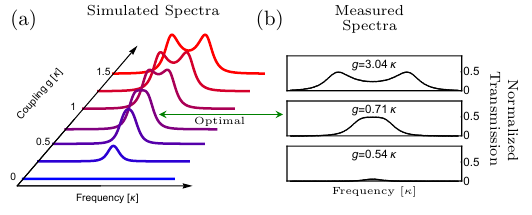}
\caption{Controlling the birefringence splitting. (a) The simulated filtered spectrum as a function of the induced birefringent coupling strength $g$. We show a full density plot of the simulated spectra in Appendix~\ref{appendix-density}. (b) Experimentally measured spectra of the second-order filter. The green arrow indicates the optimal filter performance in the simulated and measured spectra. Coupling strength $g$, obtained from fitting the transmission to \eqref{eq:BirefringenceTransmission}, is shown for each spectrum. The desired flat-transmission performance is achieved at the optimal coupling value of $g=\nicefrac{\kappa}{\sqrt{2}} \approx 0.71 \kappa$ where $\kappa$ is the linewidth of the coupled cavity system.}
\label{fig:CouplingSpectrum}
\end{figure}

\subsection{Filter Transmission Spectra}

The Hamiltonian under a birefringence splitting of $g$ is
\[\hat{H}_\mathrm{birefringence}(g) = \omega_c\mathbb{I}_2+\frac{1}{2}
\left(\begin{matrix}\label{eq:BirefringenceHamiltonian}
        i\frac{\kappa}{\sqrt{2}}&g\\g&i\frac{\kappa}{\sqrt{2}}
\end{matrix}\right),\]
and the filtered spectrum, i.e., the transmission through the other mode, is given by
\begin{align}
\begin{split}\label{eq:BirefringenceTransmission}
    T(g)=&\left|\frac{\kappa}{\sqrt{2}}\left\langle1\middle|\frac{1}{\hat{H}_\mathrm{birefringence}(g)-\omega_l\mathbb{I}_2}\middle|2\right\rangle\right|^2\\
    =&\frac{8g^2\kappa^2}{4g^4+4g^2(\kappa^2-8\delta^2)+(\kappa^2+8\delta^2)^2},
\end{split}
\end{align}
as shown in Fig.~\ref{fig:CouplingSpectrum}(a). By changing $\theta, \phi$ of the $m$-th order HWP or by changing the stress on the glass plate, we can operate at different values of g. As seen in Fig.~\ref{fig:CouplingSpectrum}(b), we obtained several filtered light spectra which exhibit the under-coupled, critically-coupled, and over-coupled regimes. The critically coupled point sits at $g=\sqrt{2}( \nicefrac{\kappa}{2})$ where we regain our Butterworth filter form of \eqref{eq:ButterworthHamiltonian} and the transmission follows the characteristic flat-top transmission and fast decay in the wings of \eqref{eq:ButterworthTransmission}. The fundamental allowed transmission at critical coupling is 100\%. The loss factors in our experiment are further discussed in section III.

\section{Experimental Performance}

\begin{figure}[!t]
\centering
\includegraphics[width=\columnwidth]{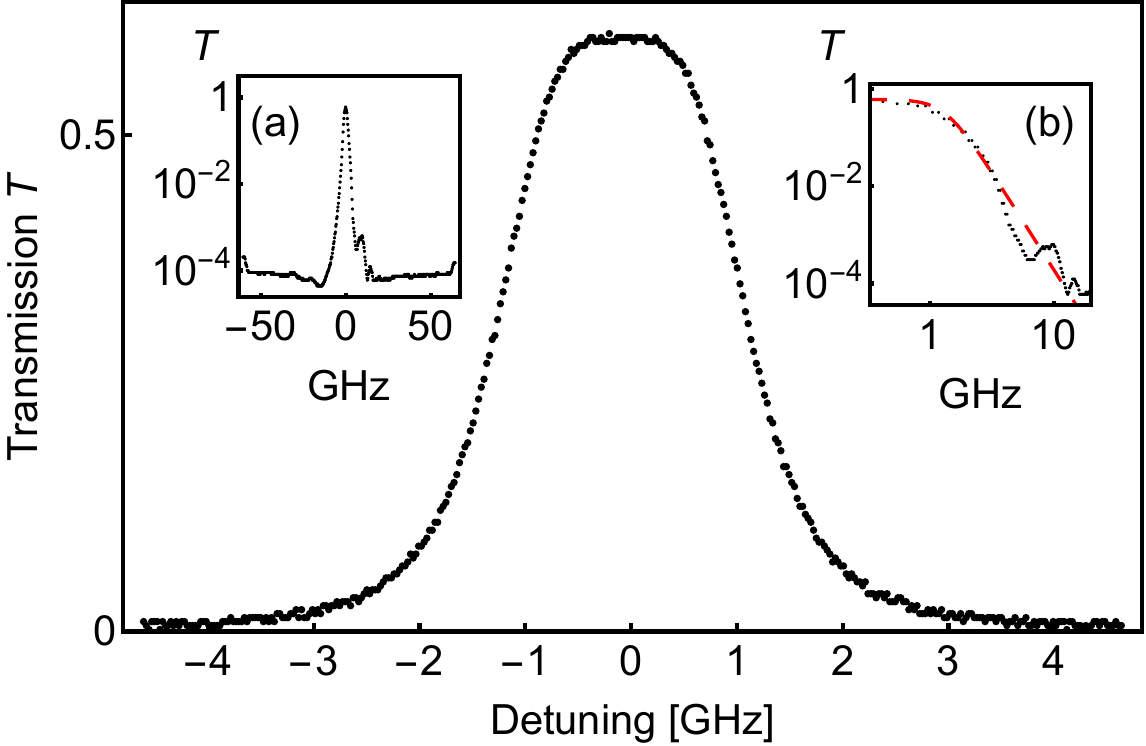}
\caption{
Filtered transmission spectrum of the second-order filter under the optimal coupling with an FSR of 125~GHz, a FWHM of $2.68(1)~\mathrm{GHz}$, and an insertion loss of $2.2(1)~\mathrm{dB}$. Transmission is normalized to input power. Inset (a): the log-linear scale plot exhibits a $\sim40~\mathrm{dB}$ suppression of background at $\nicefrac{FSR}{2}$ away from center. Inset (b): the log-log scale plot showing the fast fall-off (black points) along with the theoretically predicted fall-off (red dashed line) of a second-order filter. 
}
\label{fig:TransmissionData}
\end{figure}

With the mount shown in Fig.~\ref{fig:Pol2ndScheme}(b), we can vary the cavity length by orders of magnitude. The curved mirror and the thickness of the waveplate limit the maximum FSR to $\sim$120~GHz, while the FSR can be arbitrarily small up to the cavity stability limits imposed by the mirror curvature. For our demonstration, we assemble a second-order filter with a width of $\kappa=2\pi\times2.68(1)~\mathrm{GHz}$ ($\text{finesse}=45$). For this filter, the maximum possible suppression is $(2\times\mathrm{finesse}/\pi)^4\sim6.7\times10^5=58~\mathrm{dB}$. As shown in Fig.~\ref{fig:TransmissionData}, we have obtained an ideal second-order filter until imperfect mode-matching causes the higher-order transverse mode to appear at around 10~GHz detuning. 

The filter has an insertion loss of 2.2(1) dB, which is set by the combination of multiple loss factors, including the fiber coupling, mode matching, isolator transmission, and dielectric loss from the cavity mirror. These factors are technical limitations and can be individually and simultaneously improved upon in future iterations of this filter. The fundamental limit on the peak transmission is $100\%$.

The maximum suppression is measured to be $43~\mathrm{dB}$, bounded by the rejection limit of the IO-3-780-HP isolator rather than the resonator finesse. The maximum suppresion is seen around 15~Ghz detuning (as seen in Fig.~\ref{fig:TransmissionData}~inset~(a)) due to the combined effects of interference between different spatial modes and the insufficient polarization extinction ratio of the PBS within the isolator. In the next generation of filters, we plan to use a 50dB extinction-ratio PBS (UFPBS10A) to determine the exact scaling and better distinguish the two effects.

The behavior near resonance is well described as an ideal second-order filter, with the first deviation occurring around 3~GHz (as seen in Fig.~\ref{fig:TransmissionData}~inset~(b)), resulting from imperfect mode matching that excites higher-order transverse modes. Precise alignment allowed us to completely suppress coupling to the TEM$_{10}$ and TEM$_{01}$ modes, but suppressing higher-order modes would require further beam \emph{shaping} and, in practice, allow transmission at the  $6\times10^{-4}$ level at 10~GHz detuning. 

\section{Outlook and Applications}

We have demonstrated a minimal, robust, narrow-band high-order optical filter. We have also utilized this device to filter out the carrier and the negative sideband of an Electro-Optic Modulator (EOM) for the demonstration of a fast spatial light modulator~\cite{wei202610megahertzspatiallight}. For yet-more-stable performance, the two-mirror cavity could be replaced with a doubly-convex fused-silica etalon, with stress applied directly to the glass. Although we only described and demonstrated the second-order filter, the even ($2N$th-order) high-order filters are naturally generalizable, as shown in the Appendix. Better background rejection can be achieved by a combination of higher-performance polarizing beamsplitters and better mode matching into the resonator. A narrower passband is achievable by simply making the resonator longer. 

\section*{Acknowledgements}
Z.L. acknowledges support from the Urbanek-Chodorow Fellowship. This work was supported by the Office of Naval Research (ONR) through Grant N000142512198.

\section*{Author Contributions}
Z.L., A.V.K., and X.W. built and performed the experiments and data analysis. Z.L., A.V.K., and J.S. wrote the manuscript with contributions and input from all authors. J.S. supervised this project. Z.L., A.V.K., and X.W. contributed equally.

\section*{Competing Interests}
J.S. acts as a consultant to and holds stock options from Atom Computing.

\section*{Data Availability}
The data that support the findings of this study are available from the corresponding author upon reasonable request.

\appendix
\section{S-matrix approach for coupled cavity}

Here we derive the transmission~\eqref{eq:ButterworthTransmission} using the S-matrix approach from the first principle. 

The base of the following calculation is $2$-dimensional, with the upper one being right propagating and the lower one being left propagating. 
\begin{figure}[!htbp]
\centering
\includegraphics[height=0.3\columnwidth,page=4]{filterFig.pdf}
\includegraphics[height=0.3\columnwidth,page=5]{filterFig.pdf}
\caption{S-matrix description of the two types of coupled cavity system. }
\label{fig:Smatrix}
\end{figure}

Then, the scattering is given by
\begin{align}
\left(\begin{matrix}
u\\b
\end{matrix}\right)=S^{\mathrm{M}_1}\left(\begin{matrix}
a\\v
\end{matrix}\right),
\left(\begin{matrix}
x\\v
\end{matrix}\right)=S^{M_\mathrm{2}}\left(\begin{matrix}
u\\y
\end{matrix}\right),
\left(\begin{matrix}
d\\y
\end{matrix}\right)=S^{\mathrm{M}_3}\left(\begin{matrix}
x\\c
\end{matrix}\right),
\end{align}
where the fields are defined as in Fig.~\ref{fig:Smatrix}, and the scattering matrices are
\begin{align}
\begin{split}
S^{\mathrm{M}_1}=&\left(\begin{matrix}
ie^{ikL_1}t_{\mathrm{M}_1}&e^{2ikL_1}r_{\mathrm{M_1}}\\r_{\mathrm{M_1}}&ie^{ikL_1}t_{\mathrm{M}_1}
\end{matrix}\right),\\
S^{\mathrm{M}_2}=&\left(\begin{matrix}
it_{\mathrm{M}_2}&r_{\mathrm{M_2}}\\r_{\mathrm{M_2}}&it_{\mathrm{M}_2}
\end{matrix}\right),\\
S^{\mathrm{M}_3}=&\left(\begin{matrix}
ie^{ikL_2}t_{\mathrm{M}_3}&r_{\mathrm{M_3}}\\e^{2ikL_2}r_{\mathrm{M_3}}&ie^{ikL_2}t_{\mathrm{M}_3}
\end{matrix}\right),
\end{split}
\end{align}
where $r_{\mathrm{M}_i},t_{\mathrm{M}_i}$ are the reflectivity and transmitivity coefficients of the three mirrors.

Each of the $S$-matrix can be converted into a $\mathbf{T}$-matrix which can be calculated easily, as
\[\mathbf{T}^{\mathrm{M}_i}=\left(\begin{matrix}
S^{\mathrm{M}_i}_{11}-S^{\mathrm{M}_i}_{12}(S^{\mathrm{M}_i}_{22})^{-1}S^{\mathrm{M}_i}_{21} & S^{\mathrm{M}_i}_{12}(S^{\mathrm{M}_i}_{22})^{-1}\\
-(S^{\mathrm{M}_i}_{22})^{-1}S^{\mathrm{M}_i}_{21} & (S^{\mathrm{M}_i}_{22})^{-1}
\end{matrix}\right),\]
and 
\begin{align}
\left(\begin{matrix}
d\\c
\end{matrix}\right)=\underbrace{\mathbf{T}^{\mathrm{M}_3}\mathbf{T}^{\mathrm{M}_2}\mathbf{T}^{\mathrm{M}_1}}_{\mathbf{T}^\mathrm{tot}}\left(\begin{matrix}
a\\b
\end{matrix}\right),
\end{align}
and at $c=0$, the transmission $T$ is
\begin{align}
\begin{split}
T=&\left|\frac{d}{a}\right|^2=\left|\mathbf{T}^\mathrm{tot}_{11}-\frac{\mathbf{T}^\mathrm{tot}_{12}\mathbf{T}^\mathrm{tot}_{21}}{\mathbf{T}^\mathrm{tot}_{22}}\right|^2.
\end{split}
\end{align}

For simplicity, we let the two cavities be identical and the mirrors are lossless, i.e., $t_1=t_3,r_1=r_3, L_1=L_3=L,t_i^2+r_i^2=1$. We also introduce the detuning $\delta$ of the laser from the resonance as in the main text, so that $e^{2ikL}=e^{i\delta/\mathrm{FSR}}$, where $\mathrm{FSR}=c/2L$. This simplifies the transmission to be
\[T=\frac{t_{\mathrm{M}_1}^4t_{\mathrm{M}_2}^2}{A^2+B^2},\]
where $A=(1+r_{\mathrm{M}_1}^2)\cos(2\pi\delta\nu/\mathrm{FSR})-2r_{\mathrm{M}_1}r_{\mathrm{M}_2}$, and $B=t_{\mathrm{M}_1}^2\sin(2\pi\delta\nu/\mathrm{FSR})$. 

In the high finesse limit, $t_{\mathrm{M}_1}\ll1,t_{\mathrm{M}_2}\ll1$. We also expand the $\delta\nu$ to the $4$-th order and we obtain
\[T\sim\frac{1}{\left(\frac{4t_{\mathrm{M}_2}^2+t_{\mathrm{M}_1}^4}{4t_{\mathrm{M}_1}^2t_{\mathrm{M}_2}}\right)^2+\left(\frac{1}{2t_{\mathrm{M}_2}^2}-\frac{2}{t_{\mathrm{M}_1}^4}\right)\left(\frac{\delta}{\mathrm{FSR}}\right)^2+\frac{1}{t_{\mathrm{M}_1}^4t_{\mathrm{M}_2}^2}\left(\frac{\delta}{\mathrm{FSR}}\right)^4},\]
and is an ideal second order filter when $t_{\mathrm{M}_2}=t_{\mathrm{M}_1}^2/2$:
\[T\to\frac{1}{1+\left(\frac{2\delta}{\mathrm{FSR}t_{\mathrm{M}_1}^2\sqrt{2}}\right)^4}.\]

The intrinsic loss rate for both cavities (either formed by $\mathrm{M}_1,\mathrm{M}_2$ or $\mathrm{M}_2,\mathrm{M}_3$) is $\mathrm{FSR}/\mathrm{finesse}$, where the finesse is $1/t_{\mathrm{M}_1}^2$. As the setup, the loss rate should be $\kappa/\sqrt{2}$, i.e., $\mathrm{FSR}t_{\mathrm{M}_1}^2=\kappa/\sqrt{2}$. This leads to Eq.~\eqref{eq:ButterworthTransmission} as $T=1/(1+(2\delta/\kappa)^4)$. 

\section{2D Density Plot}
\label{appendix-density}
\begin{figure}[H]
    \centering
    \includegraphics[width=0.9\columnwidth]{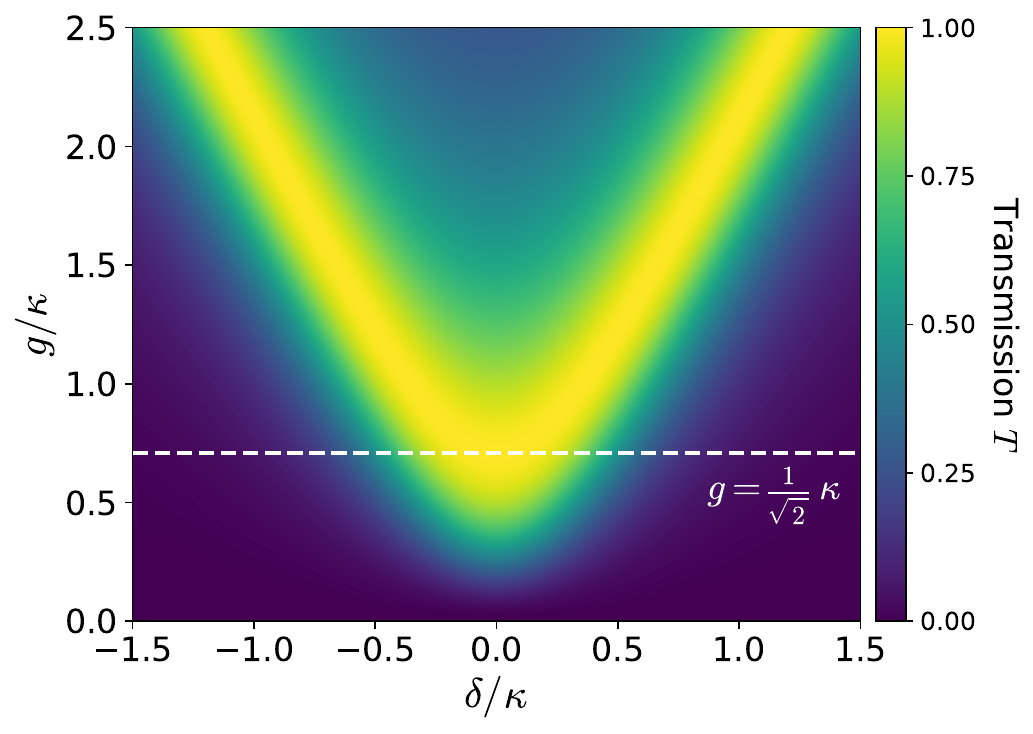}
    \caption{Density plot of the simulated
filtered spectrum. This plot depicts the form of the transmission as described by \eqref{eq:BirefringenceTransmission} as a function of the coupling strength $g$ and the detuning from the center resonance of the filter $\delta$. The white dashed line shows the critical coupling condition where $g = \nicefrac{\kappa}{\sqrt{2}}$, where the transmission behaves as in \eqref{eq:ButterworthTransmission}.}
    \label{fig:density-plot}
\end{figure}

\section{Higher-order filter}

\begin{figure}[!ht]

    \includegraphics[width=0.85\columnwidth]{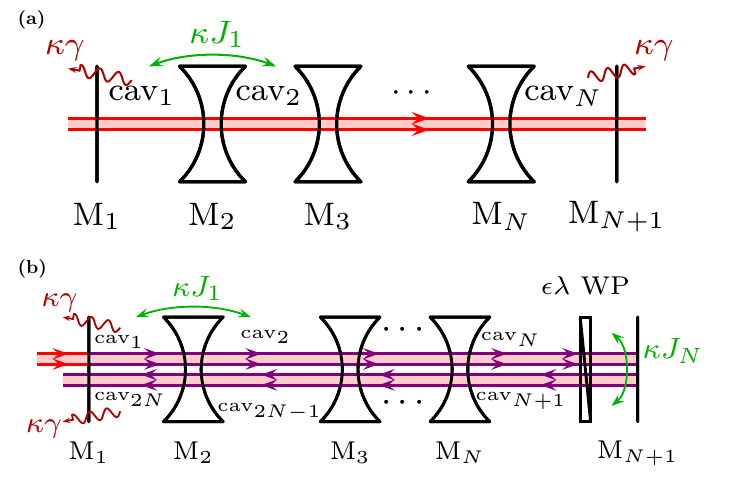}
  \caption{Higher-order filter schemes (beams are drawn as a guide for the eye). (a) A general schematic for an $N$th-order filter with $N$ coupled cavities. (b) A schematic for an even $2N$th-order filter with $N$ cavities. Each cavity has 2 polarization modes. Input light (red) is aligned to one of these polarization modes. These modes are coupled (violet) via a single intra-cavity birefringent optic to give coupled $2N$ cavity modes. The red curly arrows denote the loss of each cavity mode to the continuum and are denoted by $\kappa \gamma$. The green arrows represent the coupling between two cavity modes $n$ and $n+1$ and are denoted by $\kappa J_n$.}
  \label{fig:filters_combined}
\end{figure}

The high-order Butterworth filters can be implemented by coupling multiple cavities at the same frequency, as shown in Fig.~\ref{fig:filters_combined}(a). The specific Hamiltonian is
\begin{align}\label{eq:CouplingHamiltonian}
    \hat{H}_\mathrm{eff}=\omega_c\mathbb{I}_n+\kappa\left[\begin{array}{ccccc}
i \frac{\gamma}{2} & J_1 & 0 & \ldots & 0 \\
J_1 & 0 & J_2 & \ldots & 0 \\
0 & J_2 & 0 & \ddots & 0 \\
\vdots & \ddots & \ddots & \ddots & J_{N-1} \\
0 & \ldots & 0 & J_{N-1} & i \frac{\gamma}{2}
\end{array}\right],
\end{align}
where 
\begin{align}
\begin{split}
    J_i=&\frac{1}{4\sqrt{\sin\left[\frac{2i-1}{2N}\pi\right]\sin\left[\frac{2i+1}{2N}\pi\right]}},\\
    \gamma=&\frac{1}{2\sin\left[\frac{1}{2N}\pi\right]},
\end{split}
\end{align}
where $J_i$ are unitless tunneling rates between cavity $i$ and $i+1$ normalized to the FWHM $\kappa$, and $\gamma$ is the decay rate of the first and last cavity normalized to the FWHM $\kappa$. 
Practically, the coupling strength is determined by the mirror's transmission (assuming there is no loss)~\hbox{\cite[Ch. 12]{steck2007quantum}}
\begin{align}
g_{i,i+1}=J_i\kappa=\sqrt{\mathrm{FSR}_i\mathrm{FSR}_{i+1}}\cdot t_{\mathrm{M}_{i+1}}.
\end{align}
Specifically, the even-order filter can be efficiently implemented in the polarization basis, as shown in Fig.~\ref{fig:filters_combined}(b). Here, the coupling between cavity $N$ and $N+1$ is introduced in the same way as in the main text through a tilted multi-order HWP or stressed glass plate, while the rest of the coupling is done by the mirrors' residual transmissions. 

\bibliography{ref}

@misc{wei202610megahertzspatiallight,
      title={A 10 Megahertz Spatial Light Modulator}, 
      author={Xin Wei and Zeyang Li and Abhishek V. Karve and Adam L. Shaw and David I. Schuster and Jonathan Simon},
      year={2026},
      eprint={2601.08906},
      archivePrefix={arXiv},
      primaryClass={quant-ph},
      url={https://arxiv.org/abs/2601.08906}
}

@article{van1985multimirror,
  title={Multimirror fabry--perot interferometers},
  author={Van de Stadt, Herman and Muller, Johan M},
  journal={Journal of the Optical Society of America A},
  volume={2},
  number={8},
  pages={1363--1370},
  year={1985},
  publisher={Optical Society of America}
}

@article{Levine2018PRLHighFidelity,
  title = {High-Fidelity Control and Entanglement of Rydberg-Atom Qubits},
  author = {Levine, Harry and Keesling, Alexander and Omran, Ahmed and Bernien, Hannes and Schwartz, Sylvain and Zibrov, Alexander S. and Endres, Manuel and Greiner, Markus and Vuleti\ifmmode \acute{c}\else \'{c}\fi{}, Vladan and Lukin, Mikhail D.},
  journal = {Phys. Rev. Lett.},
  volume = {121},
  issue = {12},
  pages = {123603},
  numpages = {6},
  year = {2018},
  month = {Sep},
  publisher = {American Physical Society},
  doi = {10.1103/PhysRevLett.121.123603},
  url = {https://link.aps.org/doi/10.1103/PhysRevLett.121.123603}
}

@article{Li2022PRApplied,
  title = {Active Cancellation of Servo-Induced Noise on Stabilized Lasers via Feedforward},
  author = {Li, Lintao and Huie, William and Chen, Neville and DeMarco, Brian and Covey, Jacob P.},
  journal = {Phys. Rev. Appl.},
  volume = {18},
  issue = {6},
  pages = {064005},
  numpages = {11},
  year = {2022},
  month = {Dec},
  publisher = {American Physical Society},
  doi = {10.1103/PhysRevApplied.18.064005},
  url = {https://link.aps.org/doi/10.1103/PhysRevApplied.18.064005}
}

@article{Kimble2021PRDFilterCavityLIGO,
  title = {Conversion of conventional gravitational-wave interferometers into quantum nondemolition interferometers by modifying their input and/or output optics},
  author = {Kimble, H. J. and Levin, Yuri and Matsko, Andrey B. and Thorne, Kip S. and Vyatchanin, Sergey P.},
  journal = {Phys. Rev. D},
  volume = {65},
  issue = {2},
  pages = {022002},
  numpages = {31},
  year = {2001},
  month = {Dec},
  publisher = {American Physical Society},
  doi = {10.1103/PhysRevD.65.022002},
  url = {https://link.aps.org/doi/10.1103/PhysRevD.65.022002}
}

@article{stone2020optical,
  title={Optical mode conversion in coupled Fabry--Perot resonators},
  author={Stone, Mark and Suleymanzade, Aziza and Taneja, Lavanya and Schuster, David I and Simon, Jonathan},
  journal={Optics Letters},
  volume={46},
  number={1},
  pages={21--24},
  year={2020},
  publisher={Optical Society of America}
}

@article{nanoCoupledCavs,
author = {Xinchang Zhang and Manuj Singh and Dingning Li and Milo\v{s} A. Popovi\'{c}},
journal = {Opt. Lett.},
number = {13},
pages = {3678--3681},
title = {Engineering the passband shape of coupled-cavity filters for low loss and/or narrow bandwidth},
volume = {49},
month = {Jul},
year = {2024},
doi = {10.1364/OL.525476}
}

@article{10khzFP,
author = {C. Fabre and R. G. DeVoe and R. G. Brewer},
journal = {Opt. Lett.},
number = {6},
pages = {365--367},
title = {Ultrahigh-finesse optical cavities},
volume = {11},
month = {Jun},
year = {1986},
doi = {10.1364/OL.11.000365}
}

@article{Kessler2012-sb,
  author  = {Kessler, T. and Hagemann, C. and Grebing, C. and Legero, T. and Sterr, U. and Riehle, F. and Martin, M. J. and Chen, L. and Ye, J.},
  title   = {A sub-40-{mHz}-linewidth laser based on a silicon single-crystal optical cavity},
  journal = {Nat. Photonics},
  volume  = {6},
  number  = {10},
  pages   = {687--692},
  year    = {2012},
  month   = oct,
  doi     = {10.1038/nphoton.2012.217}
}

@article{PhysRevA.107.042611,
  title = {Sensitivity of quantum gate fidelity to laser phase and intensity noise},
  author = {Jiang, X. and Scott, J. and Friesen, Mark and Saffman, M.},
  journal = {Phys. Rev. A},
  volume = {107},
  issue = {4},
  pages = {042611},
  numpages = {23},
  year = {2023},
  month = {Apr},
  publisher = {American Physical Society},
  doi = {10.1103/PhysRevA.107.042611},
  url = {https://link.aps.org/doi/10.1103/PhysRevA.107.042611}
}

@article{PhysRevA.111.042614,
  title = {Measurement and feedforward correction of the fast phase noise of lasers},
  author = {Denecker, T. and Chew, Y. T. and Guillemant, O. and Watanabe, G. and Tomita, T. and Ohmori, K. and de L\'es\'eleuc, S.},
  journal = {Phys. Rev. A},
  volume = {111},
  issue = {4},
  pages = {042614},
  numpages = {10},
  year = {2025},
  month = {Apr},
  publisher = {American Physical Society},
  doi = {10.1103/PhysRevA.111.042614},
  url = {https://link.aps.org/doi/10.1103/PhysRevA.111.042614}
}

@article{PhysRevA.77.053809,
  title = {Subhertz linewidth diode lasers by stabilization to vibrationally and thermally compensated ultralow-expansion glass Fabry-P\'erot cavities},
  author = {Alnis, J. and Matveev, A. and Kolachevsky, N. and Udem, Th. and H\"ansch, T. W.},
  journal = {Phys. Rev. A},
  volume = {77},
  issue = {5},
  pages = {053809},
  numpages = {9},
  year = {2008},
  month = {May},
  publisher = {American Physical Society},
  doi = {10.1103/PhysRevA.77.053809},
  url = {https://link.aps.org/doi/10.1103/PhysRevA.77.053809}
}

@inproceedings{bandpassFuture,
author = {Robert L. Johnson},
booktitle = {Optical Interference Coatings 2016},
journal = {Optical Interference Coatings 2016},
pages = {TC.11},
publisher = {Optica Publishing Group},
title = {The future of high performance narrowband optical filters},
year = {2016},
doi = {10.1364/OIC.2016.TC.11}
}

@techreport{johnson2016bandpass,
  title={Bandpass Filters Past and Present},
  author={Johnson, Robert L.},
  institution={Omega Optical Inc.},
  year={2016},
  url={https://www.omegafilters.com/sites/default/files/images/Resources/Articles/Bandpass-Filters-Past-and-Present-2016HB.pdf}
}

@techreport{quarterwave2025,
  title={Quarter Wave Stack High Reflection Coatings, Interference Filters, and Bandpass Filters},
  author={MPO Technologies},
  institution={MPO Technologies},
  year={2025},
  url={https://mpo.im/tech-notes/quarter-wave-stack-high-reflection-coatings-interference-filters-and-bandpass-filters/}
}

@article{1hzFP,
author = {Mark Notcutt and Long-Sheng Ma and Jun Ye and John L. Hall},
journal = {Opt. Lett.},
number = {14},
pages = {1815--1817},
title = {Simple and compact 1-Hz laser system via an improved mounting configuration of a reference cavity},
volume = {30},
month = {Jul},
year = {2005},
doi = {10.1364/OL.30.001815}
}

@article{Hoffman:13,
author = {David P. Hoffman and David Valley and Scott R. Ellis and Mark Creelman and Richard A. Mathies},
journal = {Opt. Express},
number = {18},
pages = {21685--21692},
title = {Optimally shaped narrowband picosecond pulses for femtosecond stimulated Raman spectroscopy},
volume = {21},
month = {Sep},
year = {2013},
doi = {10.1364/OE.21.021685}
}

@article{SFG1,
author = {Lagutchev, Alexei and Hambir, Selezion A. and Dlott, Dana D.},
title = {Nonresonant Background Suppression in Broadband Vibrational Sum-Frequency Generation Spectroscopy},
journal = {The Journal of Physical Chemistry C},
volume = {111},
number = {37},
pages = {13645-13647},
year = {2007},
doi = {10.1021/jp075391j}
}

@article{SFG2,
    author = {Stiopkin, Igor V. and Jayathilake, Himali D. and Weeraman, Champika and Benderskii, Alexander V.},
    title = {Temporal effects on spectroscopic line shapes, resolution, and sensitivity of the broad-band sum frequency generation},
    journal = {The Journal of Chemical Physics},
    volume = {132},
    number = {23},
    pages = {234503},
    year = {2010},
    month = {06},
    doi = {10.1063/1.3432776}
}

@article{2dIR1,
author = {Hamm, Peter and Lim, Manho and Hochstrasser, Robin M.},
title = {Structure of the Amide I Band of Peptides Measured by Femtosecond Nonlinear-Infrared Spectroscopy},
journal = {The Journal of Physical Chemistry B},
volume = {102},
number = {31},
pages = {6123-6138},
year = {1998},
doi = {10.1021/jp9813286}
}

@article{2dIR2,
   author = "Jonas, David M.",
   title = "Two-Dimensional Femtosecond Spectroscopy", 
   journal= "Annual Review of Physical Chemistry",
   year = "2003",
   volume = "54",
   number = "Volume 54, 2003",
   pages = "425-463",
   doi = "https://doi.org/10.1146/annurev.physchem.54.011002.103907"
}

@article{EtalonRamanSpec1,
    author = {Lorenzo, Emmaline R. and Karki, Birendra and White, Katie E. and Burns, Kristen H. and Elles, Christopher G.},
    title = {Tunable FSRS measurements with reduced background signals: Using an etalon filter to generate picosecond pump pulses in the 460–650 nm range},
    journal = {The Journal of Chemical Physics},
    volume = {161},
    number = {22},
    pages = {224201},
    year = {2024},
    month = {12},
    doi = {10.1063/5.0237444},
}

@article{EtalonRamanSpec2,
title = {Super-Spectral-Resolution Raman spectroscopy using angle-tuning of a Fabry-Pérot etalon with application to diamond characterization},
journal = {Spectrochimica Acta Part A: Molecular and Biomolecular Spectroscopy},
volume = {325},
pages = {125038},
year = {2025},
doi = {https://doi.org/10.1016/j.saa.2024.125038},
author = {Yishai Amiel and Romi Nedvedski and Yaakov Mandelbaum and Yaakov R. Tischler and Hadass Tischler}
}

@article{yan2023broadband,
  title={Broadband bandpass Purcell filter for circuit quantum electrodynamics},
  author={Yan, Haoxiong and Wu, Xuntao and Lingenfelter, Andrew and Joshi, Yash J and Andersson, Gustav and Conner, Christopher R and Chou, Ming-Han and Grebel, Joel and Miller, Jacob M and Povey, Rhys G and others},
  journal={Applied Physics Letters},
  volume={123},
  number={13},
  year={2023},
  doi={10.1063/5.0161893},
  publisher={AIP Publishing}
}

@article{Hald:05,
author = {Jan Hald and Valentina Ruseva},
journal = {J. Opt. Soc. Am. B},
number = {11},
pages = {2338--2344},
title = {Efficient suppression of diode-laser phase noise by optical filtering},
volume = {22},
month = {Nov},
year = {2005},
doi = {10.1364/JOSAB.22.002338}
}

@article{Ludlow:21,
  title = {Characterization and Suppression of Background Light Shifts in an Optical Lattice Clock},
  author = {Fasano, R.J. and Chen, Y.J. and McGrew, W.F. and Brand, W.J. and Fox, R.W. and Ludlow, A.D.},
  journal = {Phys. Rev. Appl.},
  volume = {15},
  issue = {4},
  pages = {044016},
  numpages = {11},
  year = {2021},
  month = {Apr},
  doi = {10.1103/PhysRevApplied.15.044016}
}

@book{MYJbook1980,
  title = {Microwave filters, impedance-matching networks, and coupling structures},
  author = {Matthaei, George L. and Young, Leo and Jones, E. M. T.},
  publisher = {University Science Books},
  year = 1980
}

@article{Naaman2022,
  title = {Synthesis of Parametrically Coupled Networks},
  author = {Naaman, Ofer and Aumentado, Jos\'e},
  journal = {PRX Quantum},
  volume = {3},
  issue = {2},
  pages = {020201},
  numpages = {37},
  year = {2022},
  month = {May},
  publisher = {American Physical Society},
  doi = {10.1103/PRXQuantum.3.020201},
  url = {https://link.aps.org/doi/10.1103/PRXQuantum.3.020201}
}

@article{ChenEtalonShaping,
    author = {Chen, Zhao and Zhou, Xibin and Werley, Christopher A. and Nelson, Keith A.},
    title = {Generation of high power tunable multicycle teraherz pulses},
    journal = {Applied Physics Letters},
    volume = {99},
    number = {7},
    pages = {071102},
    year = {2011},
    month = {08},
    issn = {0003-6951},
    doi = {10.1063/1.3624919}
}

@article{readoutModeCleaner,
author = {N. Smith-Lefebvre and S. Ballmer and M. Evans and S. Waldman and K. Kawabe and V. Frolov and N. Mavalvala},
journal = {Opt. Lett.},
number = {22},
pages = {4365--4367},
title = {Optimal alignment sensing of a readout mode cleaner cavity},
volume = {36},
month = {Nov},
year = {2011},
doi = {10.1364/OL.36.004365}
}

@article{Levine2022PRA,
  title = {Dispersive optical systems for scalable Raman driving of hyperfine qubits},
  author = {Levine, Harry and Bluvstein, Dolev and Keesling, Alexander and Wang, Tout T. and Ebadi, Sepehr and Semeghini, Giulia and Omran, Ahmed and Greiner, Markus and Vuleti\ifmmode \acute{c}\else \'{c}\fi{}, Vladan and Lukin, Mikhail D.},
  journal = {Phys. Rev. A},
  volume = {105},
  issue = {3},
  pages = {032618},
  numpages = {13},
  year = {2022},
  month = {Mar},
  publisher = {American Physical Society},
  doi = {10.1103/PhysRevA.105.032618},
  url = {https://link.aps.org/doi/10.1103/PhysRevA.105.032618}
}

@phdthesis{simonThesis,
    author = {Simon, Jonathan},
    title = {Cavity QED with atomic ensembles},
    school = {Harvard University},
    year = {2010}
}

@BOOK{ResolventMethod,
       author = {{Cohen-Tannoudji}, Claude and {Dupont-Roc}, Jacques and {Grynberg}, Gilbert},
        title = "{Atom-Photon Interactions: Basic Processes and Applications}",
         year = {1998},
        publisher = {John Wiley \& Sons}
}

@article{BraggLaser2012,
author = {Qian Lin and Mackenzie A. Van Camp and Hao Zhang and Branislav Jelenkovi\'{c} and Vladan Vuleti\'{c}},
journal = {Opt. Lett.},
number = {11},
pages = {1989--1991},
publisher = {Optica Publishing Group},
title = {Long-external-cavity distributed Bragg reflector laser with subkilohertz intrinsic linewidth},
volume = {37},
month = {Jun},
year = {2012},
doi = {10.1364/OL.37.001989}}

@misc{steck2007quantum,
  title={Quantum and atom optics},
  author={Steck, Daniel A},
  year={2007}
}

@article{scheuer2013trap,
  title={Trap-door optical buffering using a flat-top coupled microring filter: the superluminal cavity approach},
  author={Scheuer, Jacob and Shahriar, MS},
  journal={Optics letters},
  volume={38},
  number={18},
  pages={3534--3537},
  year={2013},
  doi={10.1364/OL.38.003534},
  publisher={Optical Society of America}
}

@article{Habler2020PRA,
  title = {Higher-order exceptional points: A route for flat-top optical filters},
  author = {Habler, Nitzan and Scheuer, Jacob},
  journal = {Phys. Rev. A},
  volume = {101},
  issue = {4},
  pages = {043828},
  numpages = {11},
  year = {2020},
  month = {Apr},
  publisher = {American Physical Society},
  doi = {10.1103/PhysRevA.101.043828},
  url = {https://link.aps.org/doi/10.1103/PhysRevA.101.043828}
}

@article{Xu2019PRA,
  title = {Flat-top optical filter via the adiabatic evolution of light in an asymmetric coupler},
  author = {Xu, Xin-Biao and Guo, Xiang and Chen, Wei and Tang, Hong X. and Dong, Chun-Hua and Guo, Guang-Can and Zou, Chang-Ling},
  journal = {Phys. Rev. A},
  volume = {100},
  issue = {2},
  pages = {023809},
  numpages = {7},
  year = {2019},
  month = {Aug},
  publisher = {American Physical Society},
  doi = {10.1103/PhysRevA.100.023809},
  url = {https://link.aps.org/doi/10.1103/PhysRevA.100.023809}
}

@article{butterworth1930theory,
  title={On the theory of filter amplifiers},
  author={Butterworth, Stephen},
  journal={Wireless Engineer},
  volume={7},
  number={6},
  pages={536--541},
  year={1930}
}

\end{document}